\documentclass[aps,prl,twocolumn,showpacs,superscriptaddress,showkeys]{revtex4-1}

\usepackage{amssymb}
\usepackage{amsmath}
\usepackage{amsfonts}
\usepackage{graphicx}
\usepackage{color}
\usepackage{xspace}
\usepackage{ulem}
\usepackage{mathtools}
\usepackage{hhline}

\newcommand{\AddrCoimbra}{Univ Coimbra, Faculdade de Ci\^encias e Tecnologia da Universidade de Coimbra and CFisUC, Rua Larga, 3004-516 Coimbra, Portugal}

\begin{document}

\title{Evaporating Kerr black holes as probes of new physics}

\author{Marco Calz\`a}    \email{mc@student.uc.pt}\affiliation{\AddrCoimbra}
\author{Jo\~{a}o G.~Rosa} \email{jgrosa@uc.pt}\affiliation{\AddrCoimbra}

\date{\today}

\begin{abstract}
In the string axiverse scenario, primordial black holes (PBHs) can sustain non-negligible spin parameters as they evaporate. We show that tracking both the mass and spin evolution of a PBH in its final hour can yield a purely gravitational probe of new physics beyond the TeV scale, allowing one to determine the number of new scalars, fermions, vector bosons and spin-3/2 particles. Furthermore, we propose a multi-messenger approach to accurately measure the mass and spin of a PBH from its Hawking photon and neutrino primary emission spectra, which is independent of putative interactions between the new degrees of freedom and the Standard Model particles, as well as from the Earth-PBH distance.
\end{abstract}


\maketitle



Several cosmological scenarios predict the formation of primordial black holes (PBHs) in the early Universe, typically within a very broad mass range \cite{Hawking:1971ei, Zeldovich:1967lct,Carr:1974nx,Carr:2009jm,Carr:1975qj,Zeldovich:1967lct,Carr:2009jm,Martin:2019nuw,Carr:2020xqk,Bastero-Gil:2021fac,Garcia-Bellido:1996mdl,Bugaev:2011wy,Garcia-Bellido:2017mdw,Carr:2017edp,Germani:2017bcs,Kohri:2012yw,Kawasaki:2012wr,Bugaev:2013vba,Jedamzik:1999am,Meszaros:1975ef}. Although there is yet no concrete evidence for their existence, PBHs could be behind part of the gravitational wave signals observed by the LIGO-Virgo-Kagra detectors \cite{LIGOScientific:2016aoc} or even the ultra-short microlensing events observed by OGLE \cite{Niikura:2019kqi}. 

This means that much lighter PBHs could also have been formed, in particular PBHs born with a mass $M\simeq 5\times 10^{11}$ kg with a lifetime close to $14$ Gyrs, which would now be in the final stages of their evaporation process through Hawking emission. Bounds on the abundance of these PBHs have been obtained from their contribution to the extra-galactic $\gamma$-ray background through Hawking emission, and they can account for at most a fraction $f< 2\times 10^{-8}$ of the dark matter density \cite{Carr:2009jm, Carr:2016hva, Carr:2020gox, Arbey:2019vqx, Ferraz:2020zgi, Auffinger:2022khh,Escriva:2022duf}. Given the local dark matter density $\rho_{DM}\simeq 0.4\ \mathrm{GeV}\mathrm{cm}^{-3}$ and the Solar System's velocity in the dark matter halo $\sim 200-300$ km/s ($\sim 10^{-4}$ pc/yr), this bound translates into a PBH flux $\lesssim2\times10^5\ \mathrm{pc}^{-2}\mathrm{yr}^{-1}$, implying that every year there can up to one ``exploding'' PBH passing within a milliparsec of the Earth. This means that one may still  hope to observe PBHs in their final stages  within the lifetime of existing $\gamma$-ray telescopes. Experiments such as H.E.S.S \cite{Glicenstein:2013vha, Tavernier:2019exh}, Milagro \cite{Abdo:2014apa}, VERITAS \cite{Archambault:2017asc}, HAWC \cite{HAWC:2019wla} and Fermi-LAT \cite{Fermi-LAT:2018pfs} have already performed dedicated searches for exploding PBHs in our neighbourhood, albeit so far no detections have been reported.

Detecting PBH Hawking radiation would be extremely important for probing the dynamics of quantum fields in curved space-time, but it may also constitute a very interesting particle physics ``laboratory'', complementary to high-energy particle colliders. In \cite{Baker:2021btk, Baker:2022rkn}, in particular, it was shown that the existence of new particles beyond the TeV scale, predicted in several extensions of the Standard Model (SM) of particle physics, could significantly speed up a PBH's evaporation in its final stages, when its Hawking temperature goes beyond the electroweak scale. 

Since Hawking emission is a purely gravitational process, PBH evaporation depends only on the total number of particles with mass below the Hawking temperature, so this could yield at most limited information about the type of new particles involved, making it difficult to distinguish between different beyond the SM scenarios. 

In this Letter, we take an important step further and show that a lot more information can be extracted from PBH evaporation if one can track not only the rate at which it loses mass and but also the evolution of its spin parameter $\tilde{a}=J M_P^2/M^2$, where $J$ denotes the PBH angular momentum ($0\leq \tilde{a}<1$). 

If a PBH can only emit the SM particle content, it will quickly spin down as it evaporates, since the majority of known particles have spin and thus carry away the PBH's angular momentum, as assumed in numerous studies (e.g. \cite{Perez-Gonzalez:2020vnz,Yuan:2021xdi, Calabrese:2021src, Cheek:2021cfe, Bernal:2022pue, Bernal:2022swt, Calabrese:2022rfa, Calabrese:2023key}). Only scalar particles can be emitted as spherical waves ($l=0$), as shown by Chambers, Hiscock and Taylor \cite{Chambers:1997ai,Chambers:1997ax,Taylor:1998dk} more than two decades ago. This is, in fact, the dominant scalar emission mode at low BH spin, such that each scalar quantum emitted reduces the BH mass but conserves $J$, therefore increasing $\tilde{a}$ and the BH's angular velocity, $\Omega_H$. 

While in the SM only Higgs doublet and pion emission can spin up a PBH, which is overwhelmed by quark, lepton and gauge boson emission, this picture changes considerably in the context of string theory, the most promising framework for a quantum description of gravity alongside the other fundamental interactions. As argued in \cite{Arvanitaki:2009fg}, string theory compactifications incorporating the Peccei-Quinn solution to the strong CP problem typically predict not just one but hundreds or even thousands of light axions. These scalar fields are the four-dimensional manifestation of the higher-dimensional gauge fields appearing in the closed and open string spectrum and of the intricate geometry of the six compact dimensions. Their mass is generated only by non-perturbative effects, so that they are typically very light.

In \cite{Calza:2021czr}, we have shown with March-Russell that the existence of $N_a\gtrsim 100$ axions lighter than a few MeV (the initial temperature of PBHs born with $M\sim 10^{12}$ kg) completely changes the evolution of the PBH spin, so that even PBHs born with $\tilde{a}\lesssim 0.01$ in the radiation-dominated epoch \cite{Mirbabayi:2019uph, DeLuca:2019buf} can reach $\tilde{a}>0.1$ throughout their evolution. In fact, for $N_a\gtrsim 400$ light axion emission dominates the evaporation process and the PBH spin tends to a constant value ($\tilde{a}=0.555$ for $N_a\rightarrow \infty$ as found in \cite{Chambers:1997ai,Chambers:1997ax,Taylor:1998dk}). This offers not only an opportunity to probe the {\it string axiverse} itself by measuring the present mass and spin distribution of PBH remnants but also, as we will now show, a novel way to probe particle physics beyond the TeV scale (see also e.g.~\cite{Arbey:2019mbc, Arbey:2019jmj, Arbey:2020yzj, Hooper:2020evu, Masina:2021zpu, Arbey:2021ysg, Gregory:2021ozs, Cheek:2021odj, Cheek:2022dbx, Cheek:2022mmy} for some recent studies of the effect of spin on PBH evaporation).

The evolution of a PBH's mass and spin is determined by the dimensionless mass and angular momentum loss rates  $\mathcal{F}(\tilde{a})\equiv -(M^3/M_P^4)(\dot{M}/M)$ and $\mathcal{G}(\tilde{a})\equiv-(M^3/M_P^4)(\dot{J}/J)$, which are given by \cite{Page:1976df,Page:1976ki,Page:1977um}:
\begin{equation}\label{f_g}
\begin{pmatrix}
\mathcal{F}\\
\mathcal{G}
\end{pmatrix}=\sum_{i,l,m}\frac{1}{2\pi} \int_0^{\infty}d\bar\omega \frac{\Gamma^s_{lm}}{e^{\varpi/T_H}\pm 1}
\begin{pmatrix}
\bar\omega\\
m \tilde{a}^{-1}
\end{pmatrix}~,
\end{equation}
where the sum is taken over all particle species $i$ of spin $s$ and angular momentum quantum numbers $(l,m)$, $\varpi=\omega-m\Omega_H$, $\bar\omega=\omega M/M_P^2$ is the normalized frequency of each mode and  $T_H=\sqrt{1-\tilde{a}^2}/4\pi r_+$ is the Hawking temperature. The PBH horizon is located at $r_+=M(1+\sqrt{1-\tilde{a}^2})/M_P^2$ in natural units. The $-$ ($+$) sign corresponds to bosons (fermions). Besides the frequency integration, the non-trivial part of computing $\mathcal{F}$ and $\mathcal{G}$ resides in  determining the transmission coefficients or ``gray-body'' factors, $\Gamma^s_{lm}$. This requires numerically solving the Teukolsky master equation that governs the dynamics of spin-$s$ fields in the Kerr space-time for each mode, which we have done using a standard shooting technique (see e.g.~\cite{Press:1973zz, Rosa:2016bli,Rosa:2012uz} for details). Due to the Boltzmann suppression factors in Eq.~(\ref{f_g}), we may to a good approximation consider that each field only contributes to the mass/spin loss rates once the Hawking temperature exceeds its mass and therefore neglect the latter in computing the gray-body factors, which simplifies the numerical procedure considerably.

In our analysis \cite{foot1}, we include a massless graviton (assumed to be the only spin-2 particle), the SM degrees of freedom as $T_H$ crosses the corresponding mass thresholds, an arbitrary number $N_a$ of light axions and new particles at one or multiple mass thresholds above the electroweak scale. For the QCD degrees of freedom, we include pions for $m_\pi<T_H<\Lambda_{QCD}$ and free quarks and gluons according to the constituent masses given in \cite{Halzen:1990ip,Halzen:1991uw,MacGibbon:1991tj,MacGibbon:1991vc,MacGibbon:2007yq,Ukwatta:2009xk,MacGibbon:2010nt,MacGibbon:2015mya,MacGibbon:1990zk,Ukwatta:2015iba} (the uncertainty in the exact values of the latter having no significant impact on our results). We consider new particles of spin $s=0,\, 1/2$ and 1 typical of most SM extensions, as well as possibly a spin-3/2 gravitino arising in scenarios with spontaneously broken supersymmetry (SUSY), which is natural in the context of string compactifications. We note that in this framework spin-3/2 excitations of quarks and leptons may also be light enough to be relevant to our calculation \cite{Hassanain:2009at, foot2}. In practice we compute the contribution of generic massless spin-$s$ particles to $\mathcal{F}$ and $\mathcal{G}$, specifying the particle content at each value of $T_H$ by the number of real scalars $n_0$, the number of Weyl fermions $n_{1/2}$, the number of vector bosons $n_1$ and the number of gravitino-like particles $n_{3/2}$. For instance, the full axiverse-SM content at $T_H\simeq 200$ GeV corresponds to $(n_0,n_{1/2},n_1,n_{3/2})=(4+N_a,45,12,0)$.

\begin{figure}[h] 
\includegraphics[scale=0.52]{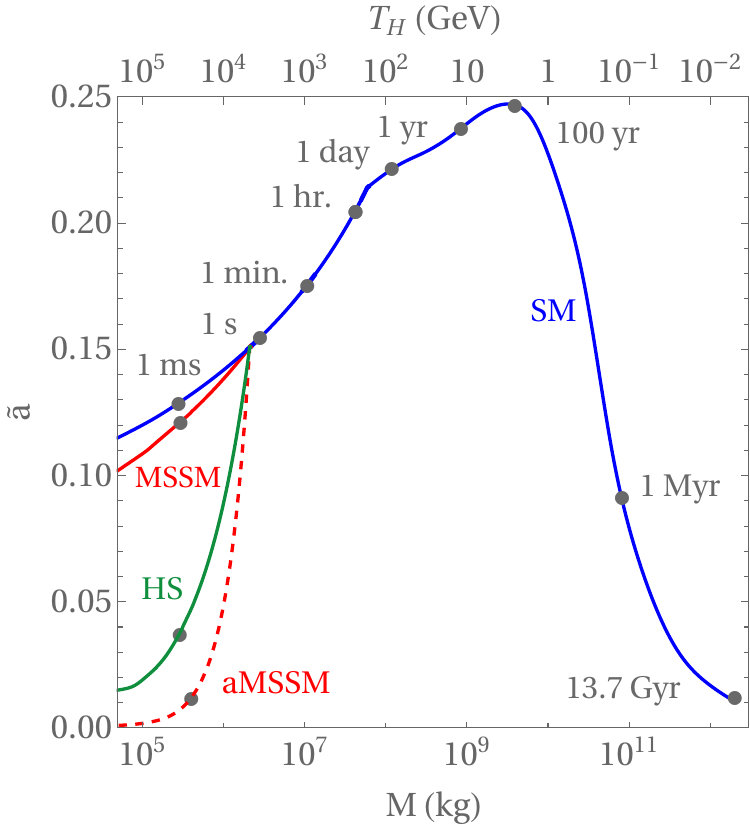}
\caption{Evolution of a PBH born with $M_i=2\times 10^{12}$ kg and $\tilde{a}_i=0.01$, considering $N_a=400$ light axions and (i) the SM particles and the graviton (blue); (ii) the MSSM with $M_{SUSY}=5$ TeV (solid red); (iii) the MSSM and axion superpartners (aMSSM) at the same SUSY breaking scale (dashed red); and a hidden sector copy of the SM (HS) also with a single mass threshold at 5 TeV (green). The labels indicate the remaining lifetime of the PBH and the upper horizontal axis gives the corresponding Hawking temperature for $\tilde{a}=0$.}
\label{fig1}
\end{figure}

The PBH evolution in the Regge plane can then be obtained by integrating:
\begin{equation}
{d\log\tilde{a}\over d\log M}={\mathcal{G}\over \mathcal{F}}-2~,
\end{equation}
across the different particle mass thresholds. As examples of the resulting evolution, we show in Fig.~\ref{fig1} the PBH evolution with $N_a=400$ light axions, the SM particles and the graviton, alongside SM extensions where all new particles have a common mass threshold at 5 TeV, namely the Minimal Supersymmetric SM (MSSM) and a hidden sector (HS) that is a copy of the SM, inspired by heterotic string scenarios with gauge group $E_8\times E_8$ (see e.g.~\cite{Berezhiani:2005ek, Chacko:2018vss, Rosa:2022sym}). These are not to be perceived as realistic scenarios, which may include multiple mass thresholds and potentially new massless particles like hidden photons or gluons, but only as simple examples that illustrate the competing effects of new particles with $s=0$ and $s\neq0$.

As one can see in this figure, the PBH spins up as a result of axion emission until its final century, when the emission of free quarks and gluons starts spinning it down once more \cite{foot3}. The spin down rate increases above the electroweak scale, mainly due to top quark emission. The new physics only has an effect in the very last second. In the MSSM example, the new scalars (squarks, sleptons and extra Higgs doublet) nearly balance the spin-down effect of the new fermions (Higgsinos, gauginos and gravitino), and the net result is only a slightly faster decrease in $\tilde{a}$ relative to the SM case. This changes drastically when one includes $N_a=400$ TeV-scale axion superpartners, saxions ($s=0$) and axinos ($s=1/2$), where in particular axino emission quickly spins down the PBH. The effect is qualitatively similar for the HS scenario, albeit not so dramatic, since the number of new fermions and gauge bosons exceeds the number of new scalars.

\begin{figure}[h] 
\includegraphics[scale=0.55]{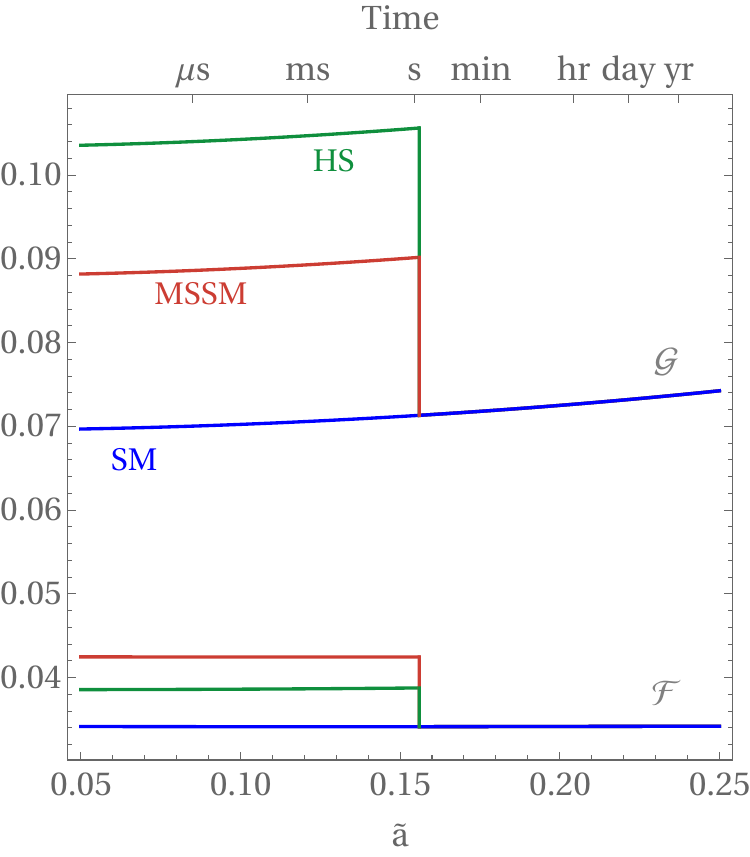}
\caption{Mass and angular momentum loss functions for the SM (blue), MSSM (red) and HS scenarios (new particles with common mass threshold of 5 TeV), considering the emission of $N_a=400$ light axions.}
\label{fig2}
\end{figure}

These examples show that new physics may have a significant impact on the PBH's Regge trajectory, reflecting the competing effects of new scalar and non-scalar particles. The MSSM and HS scenarios also illustrate that SM extensions with a comparable number of new degrees of freedom, which is $\mathcal{O}(100)$ in both cases, can lead to very distinct PBH Regge trajectories. In Fig.~\ref{fig2} we plot the mass and angular momentum loss functions for these scenarios. Both lead to sharp changes in $\mathcal{F}$ and $\mathcal{G}$ \cite{foot4} that signal new particles being emitted by the PBH, but the MSSM particles lead to a sharper increase in $\mathcal{F}$, while the HS scenario has a more pronounced effect on $\mathcal{G}$.

The loss functions $\mathcal{F}(\tilde{a})$ and $\mathcal{G}(\tilde{a})$ may in principle be determined if one can measure a PBH's mass and spin as a function of time, as we discuss in more detail below. Accurate measurements of these functions may, in fact, provide crucial information about new physics, since they are linear in the number of particles of each spin, $\mathcal{F}=\sum_s\mathcal{F}_sn_s$ and $\mathcal{G}=\sum_s\mathcal{G}_sn_s$. This means that at each stage in the PBH evaporation we may obtain two linearly independent relations between the particle numbers $(n_0, n_{1/2}, n_1, n_{3/2})$. If, in addition, one fits linear functions to the $\mathcal{F}$ and $\mathcal{G}$ data in small intervals around each value of $\tilde{a}$, we may determine their derivatives, $\mathcal{F}'(\tilde{a})$ and  $\mathcal{G}'(\tilde{a})$. These would provide two further linearly independent relations between the particle numbers, thus fully determining the particle content at each point in the Regge trajectory (assuming no new particles with spin $\geq 2$). This may then yield the number of particles in each spin representation with mass below the corresponding Hawking temperature.

Before the last second of a PBH's lifetime, this procedure can yield the number of light scalars in the spectrum, $N_a$, these being the only addition to the SM below the TeV scale that can counteract the PBH's otherwise inevitable spin down. In fact, finding even a single Kerr PBH that sustains $\tilde{a}\gtrsim 0.1$ as it evaporates would constitute a smoking-gun for a string axiverse. 

We note that, in this context, the slope of the mass loss function $\mathcal{F}(\tilde{a})$ may, however, be too small to be accurately determined, as apparent in Fig.~\ref{fig2}. This results from a partial cancellation between the contribution of scalar particles (namely the axions), since $\mathcal{F}_0'<0$ for $\tilde{a}<0.62$, and of the remaining particles, since $\mathcal{F}'_s>0$ for $s>0$. In this case, a further linear relation between the particle numbers may be found from $\mathcal{G}''(\tilde{a})$, obtained e.g.~by fitting the angular momentum loss function to a quadratic polynomial in $\tilde{a}$ within appropriate intervals.

This method could thus provide a powerful probe of new physics beyond the current reach of particle accelerators like the LHC, distinguishing different SM extensions. The PBH mass and spin could be determined from the photon primary Hawking emission spectrum if the distance to the PBH is known, which may be possible through parallax if it is sufficiently close to the Earth \cite{Calza:2021czr}. We have also recently proposed distance-independent methods to determine both Kerr parameters from the PBH spectrum, but these are either only applicable to large spin parameters $\tilde{a}>0.6$ (for which the primary spectrum exhibits a multipolar peak structure) \cite{Calza:2023gws} or rely on the spectrum of secondary photons (those radiated by e.g.~charged particles emitted by the PBH) \cite{Calza:2022ljw}. The latter depends on whether the PBH emits new charged or unstable particles that can radiate secondary photons, so cannot be employed to probe new physics in a model-independent way.

\begin{figure}[htbp] 
\includegraphics[scale=0.55]{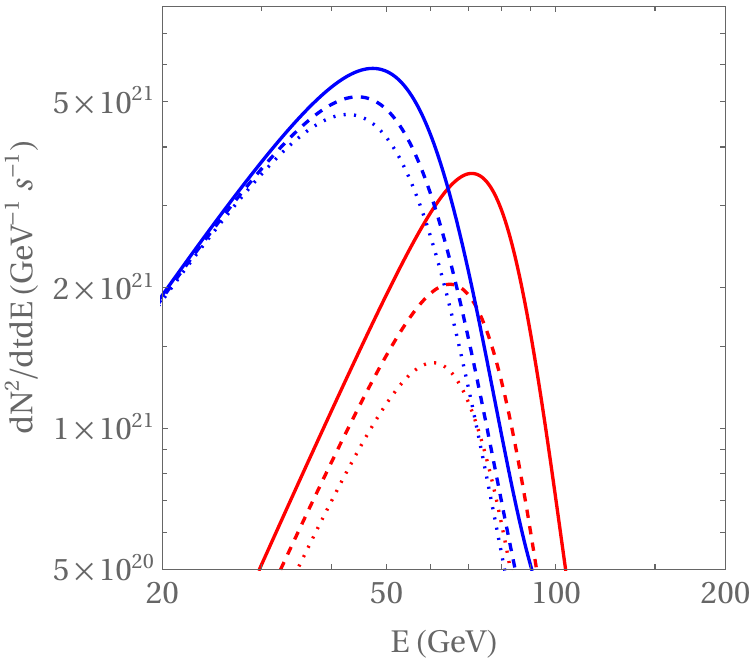}
\caption{Primary photon (red) and single neutrino (blue) Hawking emission spectra for a PBH with $M=10^{9}$ kg and $\tilde{a}=0$ (dotted), $\tilde{a}=0.2$ (dashed) and $\tilde{a}=0.5$ (solid).}
\label{fig3}
\end{figure}

Here we propose an alternative multi-messenger approach that relies on measuring the primary emission spectra of both photons and neutrinos, requiring the simultaneous detection of an evaporating PBH with $\gamma$-ray and neutrino telescopes. These spectra are given by:

\begin{equation}\label{prim}
{d^2\dot{N}_s\over dEd\Omega}={1\over 4\pi}\sum_{l,m}\!{\Gamma^s_{l,m}(\omega)\over e^{\varpi/T_H}\pm 1}\left(|{}_sS_{lm}|^2\!+\!|{}_{-s}S_{lm}|^2\right)~,
\end{equation}
where $E=\omega$ in natural units and ${}_sS_{lm}(\theta,\phi)$ are spin-weighted spheroidal harmonics. We show in Fig.~\ref{fig3} examples of the photon and (single) neutrino spectra integrated over the solid angle $\Omega$, for different values of $\tilde{a}$.

As one can see in this figure, the photon and neutrino emission peaks occur at different energies with distinct emission rates, which depend on the PBH spin. We also find that both peak energies $E_{\gamma,\nu}\propto T_H\propto M^{-1}$, while the maximum emission rates $I_{\gamma,\nu}={d\dot{N}_{\gamma,\nu}/dE} (E_{\gamma,\nu})$ are independent of the PBH mass. This means that the ratios between these quantities for neutrinos and photons depend only on $\tilde{a}$, and numerically are well fitted by:
\begin{eqnarray}\label{ratios}
    {E_\nu\over E_\gamma}=0.705 - \frac{0.559 \tilde{a}^2}{1+5.18 \tilde{a}}~,\quad
    {I_\nu\over I_\gamma}=3.423- \frac{31.05 \tilde{a}^2}{1+7.05 \tilde{a}}\,.
\end{eqnarray}
These ratios thus yield two independent ways of determining $\tilde{a}$, while the mass $M$ can then be determined from the energies $E_\gamma$ and $E_\nu$. The ratio $E_\nu/E_\gamma$ only varies by $\sim5\%$ for $0<\tilde{a}<0.5$, so that sub-percent energy resolution would be required to determine the PBH spin with $\Delta\tilde{a}\lesssim 0.1$ uncertainty, beyond the capabilities of current technology. Fortunately, the ratio $I_\nu/I_\gamma$ is much more sensitive, so that measuring it with $\mathcal{O}(10\%)$ uncertainty would yield $\Delta\tilde{a}\lesssim 10^{-2}$, which would be required to probe scenarios like the MSSM example in Fig.~\ref{fig1}.

We must, however, take into account the effects of anisotropic emission, since the measured fluxes depend on the unknown inclination of the PBH axis relative to our line of sight, $\theta$, according to Eq.~(\ref{prim}). For $0<\tilde{a}\lesssim 0.5$ both the neutrino and photon spectra are dominated by the lowest $l=m=|s|$ modes and therefore the corresponding peak emission rates are $\theta$-dependent. As proposed in \cite{Perez-Gonzalez:2023uoi}, one may determine $\theta$ by measuring the asymmetry between the neutrino and anti-neutrino fluxes (the difference between the two terms in Eq.~(\ref{prim})) or the analogous photon polarization asymmetry (although current techniques for measuring photon polarization may be difficult to extend to high-energy $\gamma$-rays). We refer the reader to the discussion in \cite{Perez-Gonzalez:2023uoi} for further details. 

An alternative way to measure the PBH inclination would be to use its proper motion, since $\theta$ will vary if the PBH is sufficiently close to the Earth, and the corresponding signal modulation may then be used to infer the integrated $I_\nu/I_\gamma$. 

One should, in fact, note that axion emission greatly reduces a PBH's lifetime, so that PBHs presently in their final stages should have been born with a larger mass (and smaller Hawking temperature) than assuming only SM particle emission, as in the example of Fig.~\ref{fig1} where the initial mass is $2\times10^{12}$ kg. This means that they would have contributed significantly less to the extra-galactic $\gamma$-ray background than in the standard scenario, relaxing the bounds on their abundance by an $\mathcal{O}(10-100)$ factor.  Therefore, we could even expect up to a few exploding PBHs crossing the Solar System within the typical lifetime of a detector.

Our proposal to determine a PBH's mass and spin does not require specially dedicated technology, since high-energy $\gamma$-ray and neutrino astronomy has a broad range of scientific goals that will surely motivate future improvements in sensitivity, energy and angular resolution of existing experiments like HAWC, Fermi-LAT, LHAASO \cite{DiSciascio:2016rgi,LHAASO:2019qtb,Cao:2023mig}  or IceCube. Several neutrino experiments like KM3Net \cite{KM3Net:2016zxf}, P-ONE \cite{P-ONE:2020ljt}, Trident \cite{Ye:2022vbk} and Baikal-GVD \cite{Baikal-GVD:2022fis} have already been proposed, alongside $\gamma$-ray telescopes with energy range beyond the TeV scale like CTA \cite{CTAConsortium:2017dvg,Gueta:2021vrf,Wild:2018cgx} and SWGO \cite{Abreu:2019ahw,LaMura:2020jzt} currently under development. We note that primary photons/neutrinos are more challenging to detect than their secondary counterparts (not shown in Fig.~\ref{fig3} for clarity), which are more numerous albeit less energetic. To give an idea of the sensitivity required, peak primary emission for a PBH at a distance of $10^{-3}$ pc corresponds to an energy flux $\sim 10^{-7}(M/10^9\ \mathrm{kg})^{-2}\ \mathrm{GeV}\mathrm{s}^{-1}\mathrm{cm}^{-2}$.

This work illustrates the enormous potential of multi-messenger astronomy to unveil new physics by accurately tracking Kerr PBH evaporation that we hope may further boost technological developments in this field. Even though at this stage we can only speculate about their existence, finding even just one of these compact objects in its last stages would allow us to probe fundamental aspects of an underlying string theory, from the expected large number of light scalars to new particles beyond the TeV scale, with PBH spin playing a key role.

\vfill


\vspace{0.5cm}
\begin{acknowledgments}
M.C. is supported by the FCT doctoral grant SFRH/BD/146700/2019. This work was supported by national funds from FCT - Funda\c{c}\~ao para a Ci\^encia e a Tecnologia, I.P., within the project UID/04564/2020 and the grant No.~CERN/FIS-PAR/0027/2021. 
\end{acknowledgments}


\begin{thebibliography}{99}
\bibitem{foot2}
We note that SM extensions could include new particles with spin $\geq 2$, e.g.~the string excitations of SM gauge bosons or Kaluza-Klein graviton modes. Their inclusion in PBH evaporation would, however, require a detailed study of Hawking emission of such particles beyond the scope of this Letter, so our analysis is restricted to SM extensions with only $s\leq 3/2$ below $\sim 100$ TeV.
\bibitem{foot4}
The loss functions should exhibit a smooth, albeit sharp, change accross particle mass thresholds when the sub-leading finite mass effects are taken into account.
\end{thebibliography}
\end{document}